\documentclass[aps,twocolumn,superscriptaddress,prl,floatfix]{revtex4}
\usepackage{amssymb}
\usepackage{amsmath}
\usepackage{graphicx}
\usepackage[left]{lineno}
\usepackage[usenames]{color}
\usepackage{url}

\def\Rq{h/2e^2}

\def\den{n_{s}}

\def\Vbg{V_{\mathrm{BG}}}
\def\Vtg{V_{\mathrm{TG}}}

\def\kF{k_{F}}

\def\nm{\mathrm{nm}}

\def\pn{\emph{p-n }}
\def\pp{\emph{p-p }}

\def\Rii{R_{ii}}
\def\Riif{R_{ii}^f}
\def\Ric{R_{ic}^f}

\def\Omt{\Omega}
\def\Ome{\Omega_{\rm{exp}}}

\begin{document}
\title{Gate-controlled guiding of electrons in graphene}
\author{J.\ R.\ Williams}
\altaffiliation{Present address: Department of Physics, Stanford
University, Stanford, CA 94305, USA.}
\affiliation{School of Engineering and Applied Sciences, Harvard University, Cambridge, MA 02138, USA}
\author{Tony\ Low}
\affiliation{School of Electrical \& Computer Engineering, Purdue University, West Lafayette, IN 47906, USA}
\author{M.\ S.\ Lundstrom}
\affiliation{School of Electrical \& Computer Engineering, Purdue University, West Lafayette, IN 47906, USA}
\author{C.\ M.\ Marcus}
\affiliation{Department of Physics, Harvard University, Cambridge, MA 02138, USA}

\date{\today}

\maketitle

{\bf Ballistic semiconductor structures have allowed the realization of optics-like phenomena in electronics, including magnetic focusing~\cite{Houten89} and lensing~\cite{Sivan90}. An extension that appears unique to graphene is to use both $n$ and $p$ carrier types to create electronic analogs of optical devices having both positive and negative indices of refraction \cite{Cheianov07}.  
Here, we use gate-controlled density with both $p$ and $n$ carrier types to  demonstrate the analog of the fiber-optic guiding in graphene \cite{Beenakker09, Zhang09, Villegas10, He10, Wu10}. Two basic effects are investigated:  (1) bipolar $\pn$ junction guiding, based on the principle of angle-selective transmission though the graphene \pn interface, and (2) unipolar fiber-optic guiding, using total internal reflection controlled by carrier density. Modulation of guiding efficiency through gating is demonstrated and compared to numerical simulations, which indicates that interface roughness limits guiding performance, with few-nanometer effective roughness extracted.  The development of \pn and fiber-optic guiding in graphene may lead to electrically reconfigurable wiring in high-mobility devices.}

\begin{figure}[b!]
\center \label{fig1}
\includegraphics[width=2.3in]{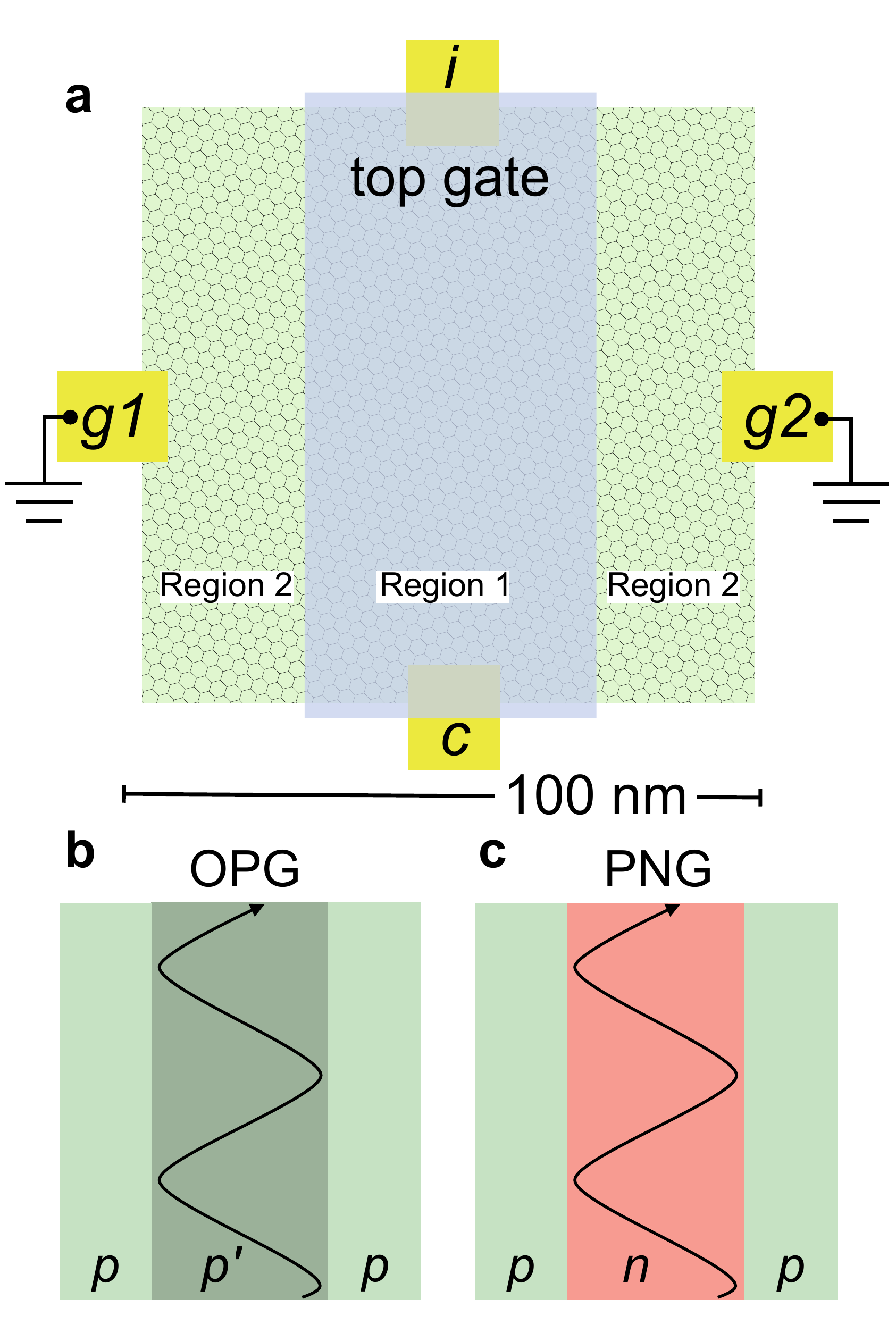}
\caption{\footnotesize{\textbf{Schematics of device and guiding in unipolar and bipolar regimes.} \textbf{a} Schematic of top-gated electron guiding device with four contacts (i, c, g1 and g2) used to measure resistances $\Rii$, $\Riif$ and $\Ric$. Voltages on top gate, $\Vtg$, and back gate, $\Vbg$, (not shown) independently control carrier densities (including sign), which serve as effective indices of refraction.  Graphene lattice orientation is schematic and is not controlled. \textbf{b}, Optical guiding (OPG) is based on reflection above a critical angle when the density in the channel (under the top gate) is higher than outside the channel (controlled by back gate), similar to the operation of a fiber optic. \textbf{c}, Alternatively, \pn guiding (PNG) is based on exponentially suppressed transmission through a \pn interface at oblique angles of incidence. }}
\end{figure}

Graphene is a single layer hexagonal lattice of carbon atoms with a gapless, linear dispersion that leads to novel electronic properties~\cite{CastroNeto09, Beenakker08}.  Carrier type and density can be controlled via gates, creating high-mobility, bipolar graphene electronics~\cite{Geim09}. Electronic transport across an interface of holes ($p$) and electrons ($n$) - a $\pn$ junction - has been studied experimentally using a combination of top/bottom electrostatic gates to create $p$ and $n$ regions~\cite{Huard07, Williams07, Ozyilmaz07}, and is now well understood theoretically~\cite{Cheianov06, Katsnelson06,Low09,Stander09, Young09}.

An intriguing possibility is to use both $n$ and $p$ carrier types in bipolar graphene structures to create electronic analogs of optical devices having both positive and negative indices of refraction. For example, a symmetrically biased $\pn$ junction has been shown theoretically to create a negative refractive index medium analogous to a Veselago lens~\cite{Cheianov07}. Such a device is not possible using conventional two-dimensional electron gas (2DEG) systems,  and demonstrates a unique feature of a Dirac-like band structure in conjunction with the ability to electrostatically tune between $n$ and $p$ carrier types. There is considerable theoretical interest in using the Dirac-like properties of graphene to create novel optical devices in graphene~\cite{Beenakker09, Zhang09, Villegas10, He10, Wu10, Park08, Low09b, Hartmann10, Mischenko10, Cesar10}, but experiments on these systems have not yet been reported.

In this Letter, we demonstrate experimentally and numerically the graphene analog of a well-known optical device, the fiber optic. Three regimes of current guiding are identified in the experiment and simulations: (1) $\pn$ junction guiding, based on the principle of angle-selective transmission, (2) the graphene fiber-optic analog, using total internal reflection and (3) an mixture of the two effects.  A metric of the guiding efficiency is predicted in all three regimes. By varying the external parameters of gate voltage and magnetic field, guiding efficiencies in each regime are extracted from experiment, where trends in the parameters are observed and compared to numerical simulations.

Photons and electrons exhibit analogous wave phenomena, reflected in the similarity of the Helmholtz equation describing electromagnetic wave propagation and the Schr\"odinger equations describing propagation for electrons~\cite{Dragoman99, Wilson93}.  In graphene, the Fermi energy ($\epsilon$) plays the role of index of refraction in an optical medium~\cite{Cheianov07, Beenakker09, Zhang09, Villegas10, He10, Wu10}, with the important feature that $\epsilon$ can be modified via electrostatic gates. We note that the dependence of the effective index of refraction on density and gate voltage in graphene differs from the dependence in conventional 2DEGs. In graphene, the wave number, and hence the effective index, is proportional to $\epsilon$. In conventional 2DEGs, the wave number and effective index scale as $\epsilon^{1/2}$~\cite{Sivan90}.  More significantly, for graphene, $\epsilon$ can either be positive (for electrons) or negative (for holes). 

For the device shown schematically in Fig.~1a, the effective index of refraction under the top gate (Region 1), $\epsilon_{1}$, is controlled by the combined voltages on the top gate, $\Vtg$, and backgate, $\Vbg$, while the effective index of refraction outside the top-gated region (Region 2), $\epsilon_{2}$, is controlled only by $\Vbg$. When $|\epsilon_{1}|>|\epsilon_{2}|$, the device operates as an electronic fiber optic with critical angle $\theta_{c}$=$\sin^{-1}(\left|\epsilon_{2}/\epsilon_{1}\right|)$, This effect, termed optical guiding (OPG), is shown schematically in Fig. 1b. Modes propagating with $\theta\geq\theta_{c}$ will be totally internally reflected and therefore travel down the channel without leaking out of the fiber.

Transmission across a graphene \pn interface decreases exponentially with angle from the interface normal~\cite{Cheianov06}. Therefore at grazing incidence, nearly all carriers impinging on the $\pn$ interface are reflected, which leads to guiding of all but a small number of carriers. This guiding mechanism is termed \pn guiding (PNG) and is shown schematically in Fig. 1c. Depending on values of $\Vtg$ and $\Vbg$, the mechanism responsible for guiding will be either OPG, $\pn$-junction guiding (PNG) or a combination of the two (OPG/PNG).  The carrier-density location of these three regions are shown as a function of the density underneath the top gate ($n_1$) and outside the top gate ($n_2$) in Fig.~2a. For unipolar devices (no $\pn$ junctions, shaded blue), only OPG is present. For bipolar devices (shaded red), both PNG and OPG/PNG can occur. OPG occurs if $|\epsilon_{1}|>|\epsilon_{2}|$; PNG occurs if $\epsilon_{1}*\epsilon_{2}<0$; OPG/PNG occurs if both conditions are satisfied.

Quantum transport simulations are used to extract guiding efficiency as a function of gate voltages (see Supp.~Info.~for numerical methods). The simulated device (Fig.~1a) has four contacts, $i$ (injector), $c$ (collector), $g1$, and $g2$ (electrical ground 1, 2) (Fig.~1a). Guiding efficiency is defined as the fraction of current collected at $c$ due to injection from $i$,  $\Omt={\cal T}_{ic}/{\cal T}_{ii}$, where ${\cal T}_{mn}$ is the transmission probability from contact $m$ to $n$, and ${\cal T}_{ii}={\cal T}_{ic}+{\cal T}_{ig1}+{\cal T}_{ig2}$. Note that $\Omega$ is finite even for equal indices of refraction in regions 1 and 2, since ${\cal T}_{ic}$ is nonzero when $\epsilon_{1}=\epsilon_{2}$ i.e. a uniformly-biased graphene sheet. We therefore define $\gamma$ as the difference between $\Omega$ and its value for equal indices, 
\begin{eqnarray}
\gamma(\epsilon_{1},\epsilon_{2})=\Omega(\epsilon_{1},\epsilon_{2})-\Omega(\epsilon_{1}, \epsilon_{1}).
\label{LB0}
\end{eqnarray}
The parameter $\gamma (\epsilon_{1},\epsilon_{2})$ then serves as an effective measure of the guiding efficiency due to unequal indices of refraction for the graphene channel, independent of the source of guiding. The condition $\Omega(\epsilon_{1}, \epsilon_{1})=\Omega(-\epsilon_{1}, -\epsilon_{1})$ follows from particle-hole symmetry. We take Fermi energy to be equivalent to the effective index of refraction, and so equate  $\epsilon_{1}$ and $\epsilon_{2}$  with Fermi energies in regions 1 and 2.

\begin{figure}[t]
\center \label{fig2}
\includegraphics[width=2.5in]{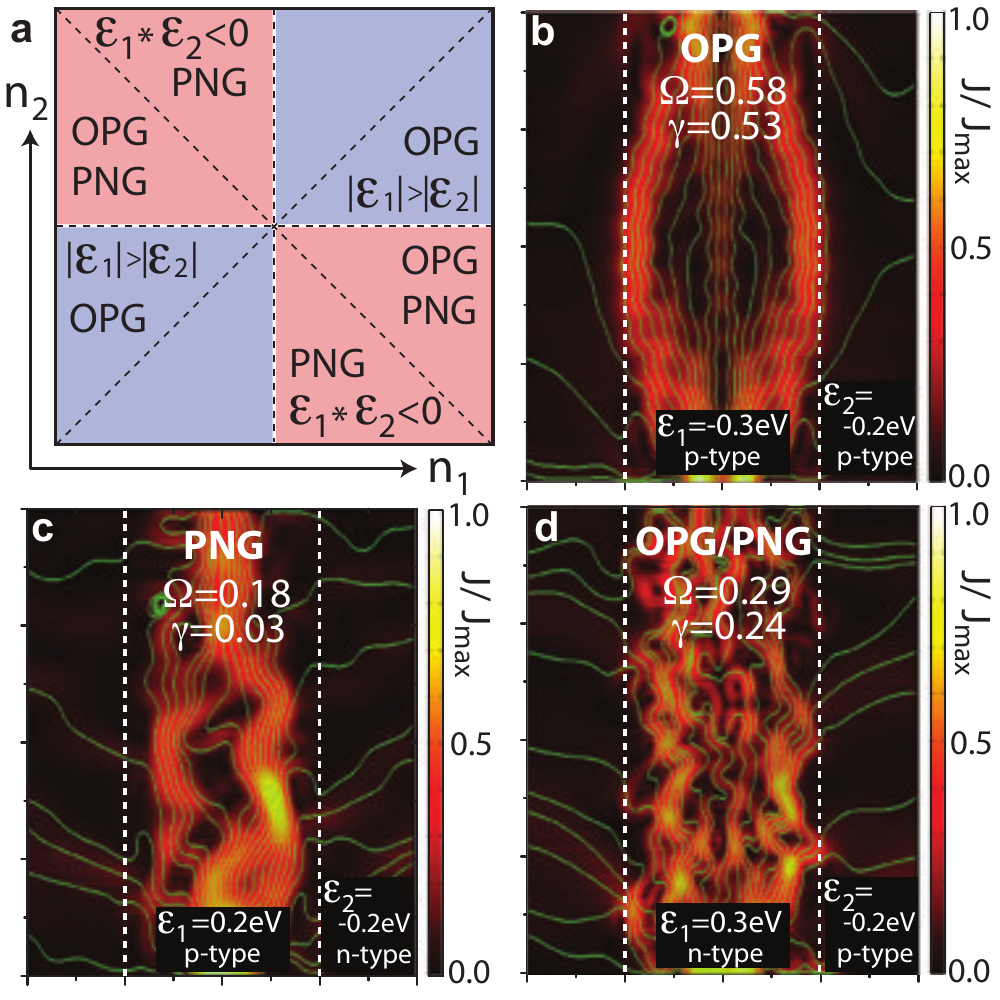}
\caption{\footnotesize{\textbf{Simulated optical and p-n guiding in gated graphene.} \textbf{a} Guiding regimes, OPG (blue), PNG and OPG/PNG (pink), as a function of density under the top gate (n$_1$) and outside the top gate (n$_2$).  The midpoint of the diagram is zero density in regions 1 and 2. \textbf{b-d}, Simulations of the current density $J$  in the three regimes for fixed effective index of refractions $\epsilon_1$ and $\epsilon_2$. The x and y-axis represent the 100\,$\nm$$\times$100\,$\nm$ size scale of the device. The guiding efficiency is $\gamma$ =0.53, 0.03, and 0.24 ($\Omega$ = 0.58, 0.18 and 0.29) in the OPG, PNG and OPG/PNG regimes, respectively. Green lines in (b-d) are lines of constant current density.}}
\end{figure}

Simulations for a 100\,$\nm$$\times$100\,$\nm$ device (matching the experimental geometry) in the OPG, PNG and OPG/PNG regimes (Fig.~2b-d), yield guiding efficiencies $\Omega(-0.3\,\mathrm{eV},-0.2\,\mathrm{eV})~=~0.58$ for OPG,  $\Omega(0.2\,\mathrm{eV},-0.2\,\mathrm{eV}) = 0.18$ for PNG, and $\Omega(0.3\,\mathrm{eV},-0.2\,\mathrm{eV}) = 0.29$ for OPG/PNG. Using $\Omega(0.3\,\mathrm{eV}, 0.3\,\mathrm{eV})$=0.05 (Fig.~2b and d) and $\Omega(0.2\,\mathrm{eV},0.2\,\mathrm{eV})$=0.15 (Fig.~2c) gives $\gamma= 0.53$ for OPG, $\gamma= 0.03$ for PNG, and $\gamma= 0.24$ for OPG/PNG. To obtain these values, simulations assumed a root-mean-square interface roughness of 4 nm in the \pn case and no roughness in the $\pp$ interface. Adding roughness is neccessary to obtain the qualitative trend observed experimentally, $\Omt^\mathrm{{OPG}}$$>$$\Omt^\mathrm{{OPG/PNG}}$$>$$\Omt^\mathrm{{PNG}}$; without interface roughness for the \pn case, numerics gave $\Omt^\mathrm{{OPG}}$$\leq$$\Omt^{\mathrm{OPG/PNG}}$.
The larger roughness in the \pn regime presumably reflecting the poor screening of disorder at the zero-density \pn junction~\cite{Zhang08} and consistent with the theoretical observation that large-angle-scattering modes are deteriorated by disorder~\cite{Rossi10}. For \pp interface, a larger $\Omt$ than experiments is obtained when assuming an ideal \pp interface. It is therefore necessary to also add some roughness to the \pp interface to obtain quantitative agreement with experiment, as seen in Fig.~3.

We next discuss the experimental realization of electron guiding. Devices were made from mechanical exfoliation of highly-oriented pyrolytic graphite.  Metallic contacts ($i$, $c$, $g1$ and $g2$) were patterned with electron-beam lithography (see Fig 1a), and the size of device was reduced to 100\,$\nm$$\times$100\,$\nm$ using an O$_2$/Ar$_2$ plasma etch, giving a device of dimensions comparable to its mean-free path, determined by transport.  A functionalized, top-gate oxide was grown \cite{Williams07} and a top-gate electrode patterned using electron-beam lithography. Differential resistance $R=dI/dV$ was measured using a standard lock-in technique at a temperature T=30\,K. Relatively high temperature and large densities were used to suppress Coloumb blockade fluctuations of the resistance present in the small device at low temperatures and densities. The field-effect, Drude mean free path $\ell_{\rm mfp}=h/e^2 \cdot \sigma/\kF$, where $\kF=\sqrt{\pi |\den|}$, for micron-sized graphene sheets is routinely found to be $\gtrsim 100~\nm$ away from the charge-neutrality point. Here, however, it is difficult to estimate $\ell_{\rm mfp}$ in these submicron devices, as the charging energy will create deviations from the simple Drude model. Extracting $\ell_{\rm mfp}$ from a four-terminal measurement $R(\Vtg)$ yields $\sim$70\,nm. Given the limitations of the Drude $\ell_{\rm mfp}$, the device is at very least in the quasi-ballistic regime. 

Experimental guiding efficiency $\Omega_{exp}={\cal T}_{ic}/{\cal T}_{ii}$ is determined from transport measurements . Transmission ${\cal T}_{ii}$ is obtained in a ``channel'' geometry, with current $I_{i}$ applied to $i$,  and $c$, $g1$, and $g2$ grounded, from the resistance $R_{ii}=V_{i}/I_{i}$,
\begin{eqnarray}
{\cal T}_{ii}\approx \frac{\Rq}{R_{ii}}.
\label{LB1}
\end{eqnarray}

Transmission ${\cal T}_{ic}$ is measured in a ``focusing" geometry~\cite{Houten89}, with current $I_{i}$ applied to $i$, voltage measured at $c$, and $g1$ and $g2$ grounded. Using this geometry, two resistances are measured by monitoring the voltages at contacts $i$ and $c$: $R_{ii}^{f}=V_{i}/I_{i}$ and $R_{ic}^{f}=V_{c}/I_{i}$. A calculation of $\Ome$ could be made by taking the ratio of these two resistances, however, the symmetry ${\cal T}_{ic}={\cal T}_{ci}$ and ${\cal T}_{ii}={\cal T}_{cc}$, may not hold in real devices. Accounting for deviation from above idealities,  we average $\Ome$ over the two configurations  $\Omega_{exp}={\cal T}_{ci}/{\cal T}_{ii}$ 
and $\Omega_{exp}={\cal T}_{ic}/{\cal T}_{cc}$ to obtain a value for guiding in terms of resistances $R_{ii}^{f}$ and $R_{ic}^{f}$ (see Supplementary information),
\begin{eqnarray}
\Ome\approx \frac{1}{2}\left( \frac{R_{ii}^{f}-R_{ii}}{R_{ic}^{f}}+\frac{R_{ic}^{f}}{R_{ii}^{f}}   \right ).
\label{LB2}
\end{eqnarray}
where the two terms in Eq. \ref{LB2} are ${\cal T}_{ci}/{\cal T}_{cc}$ and ${\cal T}_{ic}/{\cal T}_{ii}$ respectively. Ideally, the symmetry of the device would entails ${\cal T}_{ci}={\cal T}_{ic}$ and ${\cal T}_{ii}={\cal T}_{cc}$ and Eq. \ref{LB2} would reduce to $\Omt={\cal T}_{ic}/{\cal T}_{ii}$.
However, this will not be the case in typical experimental condition due to disorder and contact misalignment. A quantitative estimation of the experimental guiding efficiency 
is obtained by taking the average of $\Omega_{exp}={\cal T}_{ci}/{\cal T}_{ii}$ 
and $\Omega_{exp}={\cal T}_{ic}/{\cal T}_{cc}$, producing Eq. \ref{LB2}.
Since $0<\Ome<1$, one also has the inequalities $R_{ii}^{f}-R_{ii}>0$, $R_{ii}^{f}-R_{ii}<R_{ic}^{f}$ and $R_{ic}^{f}<R_{ii}^{f}$, in agreement with experiment (see Supplementary information).

\begin{figure}
\center \label{fig3}
\includegraphics[width=2.5in]{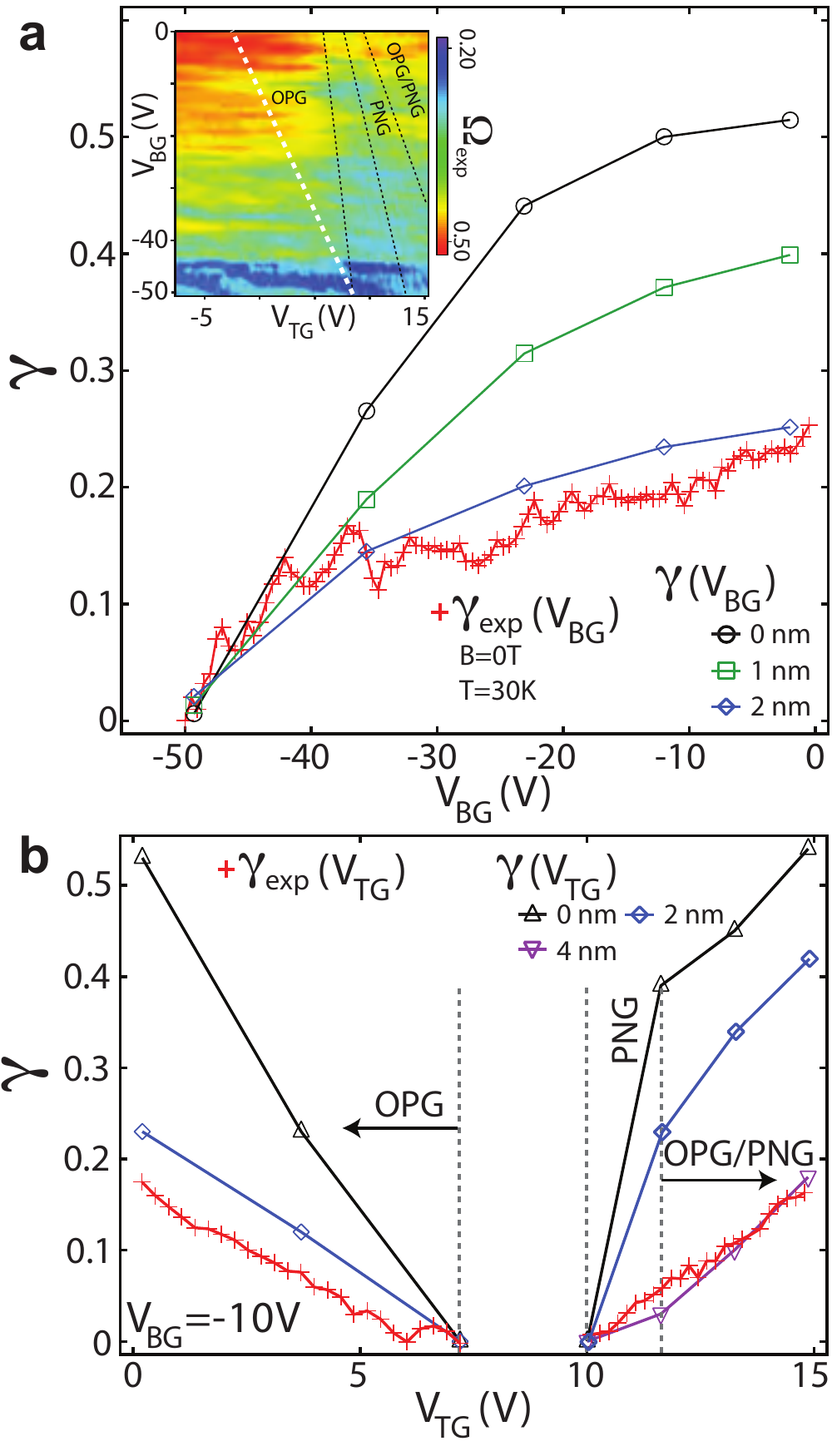}
\caption{\footnotesize{\textbf{Effects of gating and disorder on guiding efficiency.} \textbf{a}, (inset): Experimental guiding efficiency $\Ome$ (color scale) as a function of top-gate voltage, $\Vtg$, and back gate voltage, $\Vbg$, extracted from the resistances $\Rii$, $\Riif$ and $\Ric$, with three guiding regimes indicated (black dashed lines). Cut line (white dashed line) through OPG regime where experimental and numerical guiding efficiencies are compared in main figure.  (main): Experimental guiding efficiency $\gamma_{\rm exp}$, along constant fermi energy (effective index of refraction) in region 1, $\epsilon_1=0.3\,$eV. The values of $\gamma_{\rm exp}$ rises from 0 at $\Vbg=-50\,$V---the point where the density is a constant in the device---to $\sim$0.20 at $\Vbg=0\,$V. Numerical guiding efficiencies are plotted for interface roughnesses of 0 (black circles), 1\,nm (green squares) and 2\,nm (blue diamonds).  \textbf{b}, $\gamma_{\rm exp}$ extracted along $\Vbg=-10\,$V (red crosses), shows the extracted guiding in the OPG, PNG and OPG/PNG regimes along with $\gamma$ from simulation with disorder of 0 (black) and 2\,nm (blue) in the OPG regime and 0 (black), 2 (blue), and 4\,nm (purple) in the PNG and OPG/PNG regimes. Good agreement between experiment is observed for 2\,nm in the OPG regime and 4\,nm for the PNG and OPG/PNG regimes.}}
\end{figure}

Values for $\Ome$ as a function of top and back gate voltages, extracted $\Rii$, $\Riif$ and $\Ric$, are shown in the inset of Fig.~3, along with boundaries of the guiding regimes based on the boundaries in Fig.~2a. $\Ome$ is maximal in the OPG and OPG/PNG regimes, and follows the pattern $\Ome^\mathrm{{OPG}}$$>$$\Ome^\mathrm{{OPG/PNG}}$$>$$\Ome^\mathrm{{PNG}}$.

Experimental and numerical results for guiding efficiency are compared in the OPG regime, where guiding is most efficient, along a cut at constant density (Fermi energy $\epsilon_1$$\sim$ 0.3\,eV)  in Region 1, indicated by the white dashed line in the inset of Fig.~3. Experimental guiding efficiency $\gamma_{\rm exp}$ along this cut is obtained from $\Ome$ by subtracting $\Ome($0.3\,eV, 0.3\,eV)=0.26. Figure 3 shows $\gamma_{\rm exp}$ as a function of $\Vbg$ in the OPG regime along with numerical results for $\epsilon_1=0.3$\,eV with numerical interface roughnesses of 0, 1, and 2\,nm. Numerical values of $\gamma$ were computed from $\Omt$ using $\Omt($0.3\,eV, 0.3\,eV)=0.05. For all values of interface roughness, $\gamma$ increases with increasing $\Vbg$ (or decreasing $|\epsilon_{2}|$), and good quantitative  agreement with experiment is found for a roughness of 2\,nm. The dependence of $\gamma$ on $\Vbg$ can be understood by analogy with optical fiber, where decreasing the refractive index in the fiber cladding  $|\epsilon_2|$ leads to smaller critical angle. The inset of Fig.~3 shows that the trend of increasing $\Ome$ with back-gate voltage is more prominent in the OPG regime than in the PNG regime. This is consistent with expectation, as the effect of $\Vbg$ in the PNG regime is mostly to change the location of the \pn interface, where reflection occurs; transparency of the \pn interface itself is not strongly affected by $\Vbg$, as it is in the OPG regime. 

$\gamma_{exp}$ is also extracted at $\Vbg$=-10\,V, showing guiding in the OPG, PNG and OPG/PNG regimes (red crosses in Fig. 3b). Values for the ``equal-episilon'' background subtraction were extracted experimentally for the OPG regime, while the PNG and OPG/PNG values were inferred from the extracted value using particle-hole symmetry (i.e. $\Omega(\epsilon_{1}, \epsilon_{1})=\Omega(-\epsilon_{1}, -\epsilon_{1})$) as in Fig. 2. In the OPG regime, $\gamma_{exp}$ falls roughly linearly as a fucntion of $\Vtg$ from 0.18 to 0, while in the PNG and OPG/PNG regimes it rises from 0.01 to 0.16. For $\Vtg$ between $\sim$7\,V and 10\,V, $\gamma_{exp}$ in not define as it does not satisfy the conditions for guiding (see Fig. 2a). Numerical simulations for this value of $\Vbg$ are shown for the 3 regimes for 3 different amounts of disorder: 0 (black), 2 (blue) and 4\,nm (purple). Here we see good agreement between $\gamma_{exp}$ and $\gamma$ in the OPG for 2\,nm of disorder, similar to Fig. 3a. However 2\,nm of interface disorder is clearly off in the PNG and OPG/PNG regimes, and more disorder (4\,nm) is needed to obtain agreement between experiment and simulations. This is consistent with the argument of poor screening at the $\pn$ interface used in the PNG and OPG/PNG regimes in Fig. 2. 

\begin{figure}
\center \label{fig4}
\includegraphics[width=2.3in]{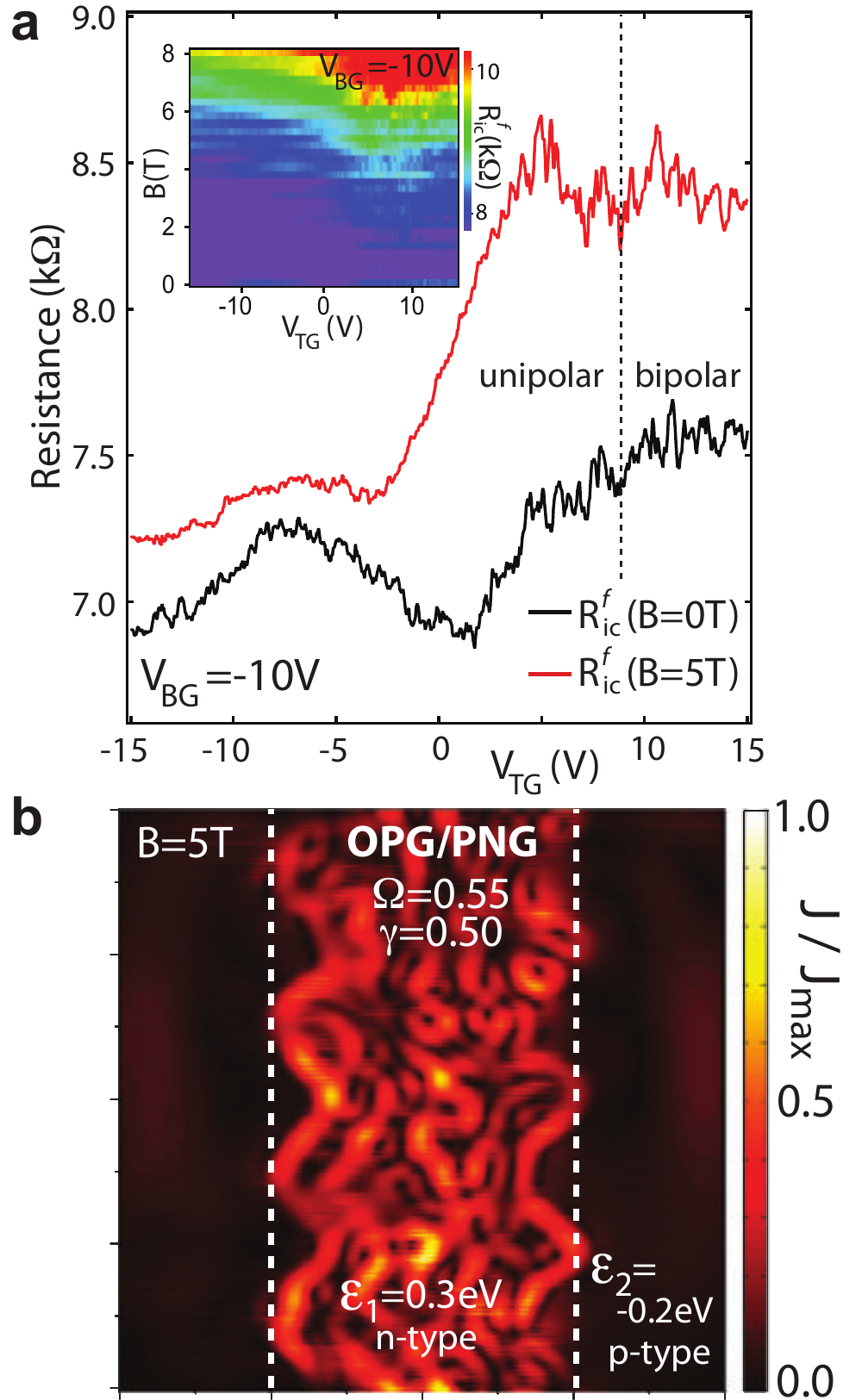}
\caption{\footnotesize{\textbf{Magnetic field improves gate-defined guiding.} \textbf{a}, (inset) $\Ric$ as a function of $\Vtg$ and B at $\Vbg$=-10\,V. (main) Plot of $\Ric$ at B=0\,T and B=5\,T. The increase in B corresponds to an increase in $\Ric$ and the ratio $\Ric(\mathrm{bipolar})/\Ric(\mathrm{unipolar})$, indicating an magnetic-field enhancement of PNG contribution to $\Ric$. Inset: $\Ric(\Vtg$,B) demonstrates enhancement of $\Ric$ is evident for fields B $>$2\,T. \textbf{b}, Simulation of J in the OPG/PNG regime at B=5\,T. $\gamma$ is enhanced from 0.24 (Fig.~2d) to 0.50 ($\Omt$ is enhanced from 0.29 to 0.55).}}
\end{figure}

Transparency across a graphene \pn junction decreases with applied perpendicular magnetic field (B), as discussed theoretically by Refs.~\cite{Cheianov06, Shytov09}, and seen experimentally in Ref. \cite{Huard07}. The reduced transparency in a magnetic field increases guiding efficiency, as demonstrated experimentally in Fig.~4. When $\Ric$$^2 \gg$ $\Riif (\Riif - \Rii )$--- the experimentally relevant case, see Fig.~S1 in the Supp. Info---$\Ome \sim \Ric$. $\Ric$($\Vtg$) at B=5\,T is compared to the zero-field value in Fig.~4a. At $\Vbg$=-10\,V an increase of $\sim$0.1\,k$\Omega$ in the unipolar regime and $\sim$1\,k$\Omega$ in the bipolar regime as B is increase to 5\,T is observed. The inset of Fig.~4a shows $\Ric(\Vtg$,B), where an enhancement in resistance is apparent for B$>$2\,T. Since the ratio $\Ric(\mathrm{bipolar})/\Ric(\mathrm{unipolar})$ increases with B, we ascribe the enhancement in $\Ric$ (and, by inference, an increase in $\Ome$) as a result of an increase in the PNG contribution to current guiding. Fig.~4b shows the simulated current density with B=5\,T, showing an improved guiding efficiency of $\gamma=0.50$ ($\Omt=0.55$) from the B=0 value of $\gamma=0.24$ ($\Omt=0.29$)  [see Fig.~2d for B=0 value]. 

In summary, we have investigated electron guiding in graphene by tuning the carrier type and density using local electrostatic fields to create the analogue of an optical fiber. Guiding efficiency was extracted in three regimes: in the unipolar OPG regime, where the device is analogous to a fiber optic; in the bipolar PNG regime, where guiding occurs because of reflection at \pn interfaces; and the bipolar OPG/PNG regime, where both mechanisms operate. We also demonstrated experimentally that guiding efficiency increases with an applied perpendicular magnetic field, consistent with numerical results.  One would expect that pseudomagnetic fields created by strain engineering \cite{Guinea09} of greater than 300\,T~\cite{Levy10} can also enhance guiding. Improvements to guiding efficiency will result from reducing interface disorder. Engineering of a collimated source with modes within the acceptance cone of the fiber would also improves guiding. With such improvements, this approach could perhaps lead to electrically reconfigurable wiring using graphene.

Device fabrication was done using Harvard's Center for Nanoscale Systems (CNS), a member of the National Nanotechnology Infrastructure Network (NNIN) under NSF award ECS-0335765, and was supported in part by INDEX, an NRI Center, and the Harvard NSEC.

\end{document}